\documentclass[a4paper,10pt,]{article}
\usepackage[latin1]{inputenc}
\usepackage[fleqn]{amsmath}
\usepackage{graphicx}
\usepackage{epsfig}
\usepackage{float}
\usepackage{caption}
\usepackage{amsfonts}
\begin{document}

\begin{center}
{\Large\bf Phase space analysis of a holographic dark energy model}
\\[15mm]
Nandan Roy \footnote{E-mail: nandan@iiserkol.ac.in} and
Narayan Banerjee \footnote{E-mail: narayan@iiserkol.ac.in}

{\em Department of Physical Sciences,~~\\Indian Institute of Science Education and Research Kolkata,~~\\Mohanpur-741246,India.}\\[15mm]
\end{center}

\begin{abstract}
The stability of interacting holographic dark energy model is discussed. It is found that for some class of the rate of interaction between dark matter and dark energy, the system has a natural solution where the universe had been decelerating in the begnning but finally settles down to an acelerated phase of expansion.
\end{abstract}


\section{Introduction:}
The holographic principle ensures that the degrees of freedom of any system is determined by the the area of the boundary and not really by the volume\cite{'thooft, susskind}. Consideration of black hole thermodyamics indicates that the holographic principle actually requires a large wavelength cut-off. A very brief but comprehensive review of this infra-red cut-off and its application in cosmology as a ``holographic'' dark energy is given by Pavon\cite{diego}. Cosmological implications of this holographic dark energy, particularly its role in driving the accelerated expansion of the universe have been quite thoroughly discussed\cite{li, diego1, diego2}. Holographic dark energy has been discussed in nonminimally couple theoreis as well, such in Brans-Dicke theory by Banerjee and Pavon\cite{diego3}, and in a chameleon scalar field model by Setare and Jamil\cite{setare}. This list is not exhaustive by any means and intends to give some examples. \\

The holographic dark energy attracted attention as it can alleviate, if not resolve, the issue of cosmic coincidence, i.e., why the energy densities due the dark matter and the dark energy should have a constant ratio for the present universe\cite{diego2}. \\

The present work deals with the stability of a holographic dark energy model in standard Einstein's gravity. The model is quite general as it allows the dark energy to interact with the dark matter, so that one can grow at the expense of the other.  The method taken up is the dynamical systems study. The field equations are written as an autonomous system and the fixed points are looked for. A stable fixed point, an ``attractor'', is apt to describe the final state of the universe. This kind of dynamical systems study is not new in cosmology. There are excellent reviews on this topic\cite{ellis, coley}. However, such analysis is more frequently used where a scalar field is involved. For an inflaton field, the dynamical systems study has been used by Gunzig et al\cite{gunzig} and Carot and Collinge\cite{carot} while Urena-Lopez\cite{urena}, Roy and Banerjee\cite{nandan2} did that for a quintessence field. Kumar, Panda and Sen\cite{anjan}, Sen, Sen and Sami\cite{soma}, Roy and 
 Banerjee\cite{nandan1} and Fang et at \cite{fang} utilized the dynamical systems analysis for an axionic quintessence, a thawing dark energy, tracking quintessence and phantom, tachyonic and k-essence fields respectively. A few dynamical systems study on holographic dark energy models, such as one where the infra-red cut-off given by the Ricci length\cite{nairi1} and by the future event horizon\cite{nairi2} are there in the literature. Setare and Vagenas\cite{setare1} discussed the bounds on the effective equation of state parameters on the basis of dynamical systems study, with an infra-red cut off given by the future event horizon. The present deals with a holographic dark energy model where the cut-off is determeined by the Hubble length\cite{diego, diego2}.\\

In the second section we actually set up the autonomous system and investigate the stability of a holographic dark energy. In section 3, we discuss the bifurcation present in the system and in the last section we discuss the results.

 \section{Phase space analysis of the model:}

We consider an interacting holographic dark energy model in which the universe is filled with a pressureless matter component with energy density $\rho_{m}$ and a holographic dark energy of density $\rho_h$. There is an interaction between these two components. The total energy density of the universe is $\rho = \rho_m + \rho_h$.

In a spatially flat FRW universe with the line element
\begin{equation} \label{metric}
ds^2 = - dt^2 + a^2(t) (dr^2 + r^2 d \omega^2),
\end{equation}
Einstein's field equations are
\begin{equation} \label{field1}
3 H^2 = 8 \pi G (\rho_{m} + \rho_{h}),
\end{equation}
\begin{equation} \label{field2}
\dot{H} = - \frac{3}{2} H^2 (1 + \frac{w}{1+r}),
\end{equation}
where $w = \frac{p_{h}}{\rho_{h}}$, is the equation of state parameter of holographic dark energy, $p_{h}$ is the contribution to the pressure by the holographic dark energy and $r = \frac{\rho_{m}}{\rho_{h}}$, is the ratio of the two energy densities\cite{li, diego}.

The conservation equations for $\rho_{m}$ and $\rho_{h}$ are
\begin{equation} \label{matter}
\dot{\rho_{m}} + 3 H \rho_{m} = Q,
\end{equation}
and
\begin{equation} \label{holo}
\dot{\rho_{h}} + 3 H (1+ w) \rho_{h} = -Q,
\end{equation}
respectively. The dark matter and the dark energy are assumed to interact amongst themselves and hence do not conserve separately. They conserve together and $Q$ is the rate of loss of one and hence the gain of the other and is assumed to be proportional to the dark energy density given by $Q= \Gamma \rho_{h}$, where $\Gamma$ is the decay rate\cite{diego1}. \\

Using (\ref{matter}) and (\ref{holo}) , one can write time evolution of $r$ as 
\begin{equation} \label{r}
\dot{r} = 3 H r (1+r) [\frac{w}{1+r} + \frac{Q}{3 H \rho_{m}}].
\end{equation}

If the holographic bound is saturated, one has\cite{li}
\begin{equation} \label{bound}
\rho_{h} = 3 M_{p}^2 C^2 / L^2,
\end{equation}  
where $L$ is the infra-red cut-off that sets the holographic bound. If the cut-off is chosen to be the Hubble length, which has a clue towards the resolution of the coincidence problem\cite{diego1}, one has

\begin{equation} \label{holodensity}
\rho_{h} = 3 C^2 M_{p}^2 H^2.
\end{equation}

By differentiating equation (\ref{holodensity}) and using (\ref{field2}) in the result, one can write

\begin{equation} \label{rhohdot}
\dot{\rho_{h}} = -3 H (1 + \frac{w}{1+r}) \rho_{h}.
\end{equation}

Equations (\ref{holo}) and (\ref{rhohdot}) yield the expression for the equation of state parameter $w$ for the  holographic dark energy as,
\begin{equation} \label{w}
w = - (\frac{1+r}{r}) \frac{\Gamma}{3 H}.
\end{equation}
\\
From equations (\ref{bound}) and (\ref{field1}), one has $ \rho_{m} = 3 M_p ^2 H^2 (1-C^2)$. Considering a saturation of the holographic dark energy, equations (\ref{field2}) and (\ref{holo}) now yield
\begin{equation} \label{dot1}
\dot{H} = - \frac{3}{2} H^2 (1 - \frac{C^2}{3 (1- C^2)} \frac{\Gamma}{H}),
\end{equation}
and,
\begin{equation} \label{dot2}
\dot{\rho_{m}} + 3 H \rho_{m} = 3 C^2 M_p ^2 H^2 \Gamma.
\end{equation}
In the subsequent discussion, equations (\ref{dot1}) and (\ref{dot2}) will replace the field equations (\ref{field1}) and (\ref{field2}). For the study of the phase space behaviour of the system,  we introduce a new set of variables $x = \rho_{m}$ , $y = \frac{\Gamma}{H}$ and $N = \ln a$.  The system of equations can now be written in the form of an autonomous system in terms of new dynamical variables as 
\begin{equation} \label{x}
x^{\prime} = -3 x + \frac{C^2}{1-C^2} x y,
\end{equation}
\begin{equation} \label{y}
y^{\prime} = - \frac{3}{2} (\lambda - 1) y (1- \frac{C^2}{3 (1- C^2)} y).
\end{equation}
Here $\lambda = (\dfrac{d\Gamma}{dH}) / (\frac{\Gamma}{H})$ and `prime' indicates a differentiation with respect to $N$. In an FRW cosmology, $H$, the fractional rate of change of the length scale of the universe, is the naturally available rate. We assume the decay rate $\Gamma$ to be a function of $H$, $\Gamma = \Gamma(H)$. Depending on the value of $\lambda$, we have classified our system into two classes. \\
\\
I) When $\lambda \neq 1$, $\Gamma$ is any function of $H$, except a linear function.\\
\\
II) When $\lambda =1 $, $\Gamma$ is linear functions of H. \\
\\
In order to discuss the phase space behaviour of a dynamical system, written in the form 
\begin{equation*}
z_{i}^{\prime} = f_{i}(z_{j}), \hspace{.6cm} i,j=1,2,.....n,
\end{equation*}
one has to find the fixed points of the system. Here $z_{i}^{\prime} = \dfrac{dz_{i}}{dt}, z_i = {z_1,....z_n} \in \mathbb{R}^n$ and $f_i: \mathbb{R}^n \longrightarrow \mathbb{R}^n$. Fixed points ($z_i=z_i^*$) of the system are the simultaneous solutions of the equations $z_i^{\prime} = 0$ for all $i$. The stability of a fixed point can be determined from eigen values of the Jacobian matrix ($\dfrac{\delta f_i}{\delta z_j} \mid_{z_j=z_j^*}$) at that fixed point. If real part of the eigen values are negative then the fixed point is stable otherwise it is either unstable or a saddle. The details may be found in any standard text, such as in reference \cite{strog}.
\subsection{Class I: $\lambda \neq 1$}
In this case the problem is a two-dimensional one and equations (\ref{x}) and (\ref{y}) form our system. Fixed points of the system are the simultaneous solutions of the equation $x^{\prime} = 0$ and $y^{\prime} = 0$. So it is easy to check that it admits two fixed points, namely, $p_1 : (x=0, y=0)$ and $p_2 : (x$ is arbitray, $y = \frac{3 (1-C^2)}{C^2})$. The second fixed point is a set of non isolated fixed points, a straight line parallel to $x$ - axis. If Jacobian matrix at any point of a set of non isolated fixed points has at least one zero eigen value the set of fixed points is called normally hyperbolic\cite{strog}. Stability of a normally hyperbolic\cite{reza} set of fixed points can be analysed from sign of remaining eigen values. If remaining eigen values are negative then the fixed point is stable. Eigen values and stability condition of the fixed points are given in the following table.\\
\begin{center}
 Table 1
\end{center}

\begin{tabular}{|c|c|c|c|}
\hline 
Fixed Points & Co-ordinate & eigen values &  Condition of stability \\ 
\hline 
$p_1$ & $x=0,y=0$ & $-3, - \frac{3}{2} (\lambda -1)$  & $\lambda > 1$ \\ 
\hline 
$p_2$ & $x(N), y = \frac{3 (1-C^2)}{C^2} $ & $0,  \frac{3}{2} (\lambda -1)$  & $\lambda < 1$ \\ 
\hline 
\end{tabular}   
~

~
\linebreak
 Phase plots of this system has been shown in figure 1 and figure 2 for $\lambda>1$ and $\lambda < 1$ respectively. The plots show that fixed points $p_1$ and $p_2$ are indeed stable for $\lambda>1$ and $\lambda<1$ respectively. This is consistent with the fact that negative eigenvalues indicate stable fixed points (see table 1)\\

Form the field equations and the definition of $x$ and $y$, the deceleration parameter can be written as
\begin{equation}\label{q}
 q = -1 + \frac{3}{2} (1- \frac{C^2}{3(1-C^2)} y).
\end{equation}

As we have two fixed points, the system may be treated as a heteroclinic one where solutions join two fixed points.\\

For $\lambda>1$, $p_1$ is a stable fixed point and $p_2$ is an unstable one, thus indicating a sink and source respectively. So the universe can originate from  $p_2$ with an acceleration ($q=-1$) and an arbitrary $\rho_{m}$ and can settle down to $p_1$, the stable fixed point where the expansion is decelerated ($q=\frac{1}{2}$) with $\rho_{m} \longrightarrow 0$. This squarely contradicts the observation which indicates an exactly opposite situation!

For $\lambda<1$, however, the fixed points actually reverese their roles as the source and the sink. In this case $p_1$ is unstable, thus, for a small perturbation,  the universe starts evolving with a deceleration ($q=\frac{1}{2}$) and settles down to the final configuration of an accelerated expansion ($q=-1$). The final  $\rho_{m}$ is an arbitrary function of $N$. This situation is indeed realistic. 

\begin{figure}[ht]
\centering
\begin{minipage}[b]{0.45\linewidth}
\includegraphics[scale=.5]{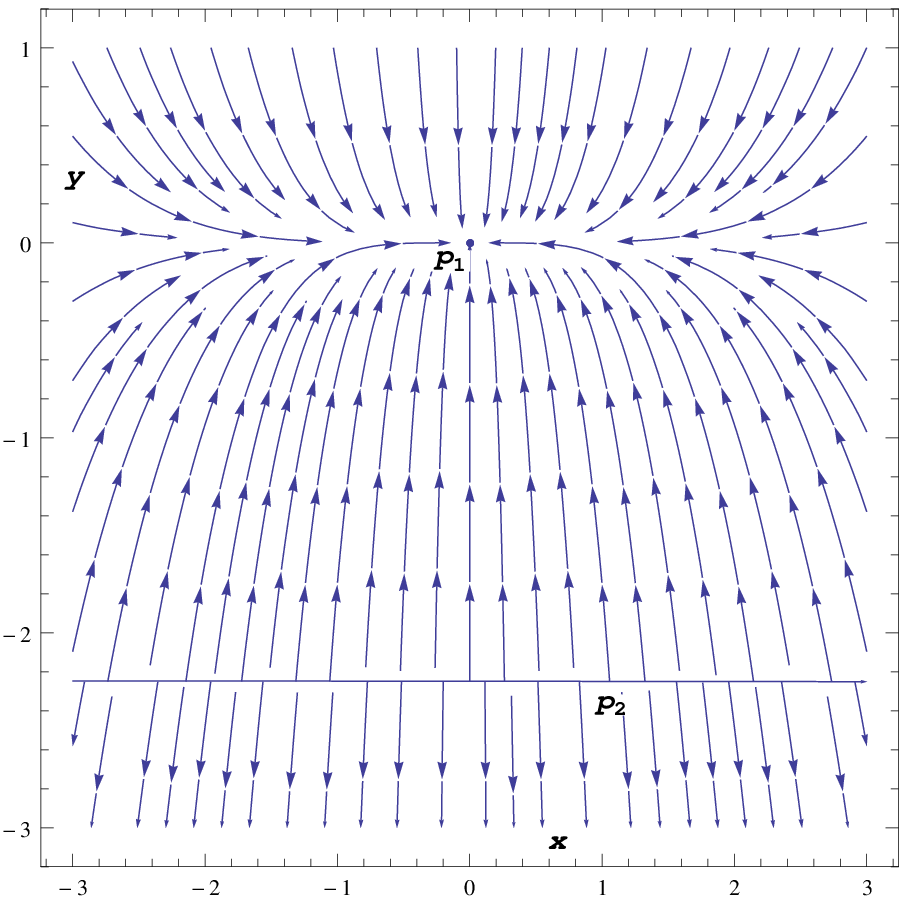}
\caption{Phase plot of the system when $\lambda = 10$ and $C=2$.}
\label{fig:minipage1}
\end{minipage}
\quad
\begin{minipage}[b]{0.45\linewidth}
\includegraphics[scale=.5]{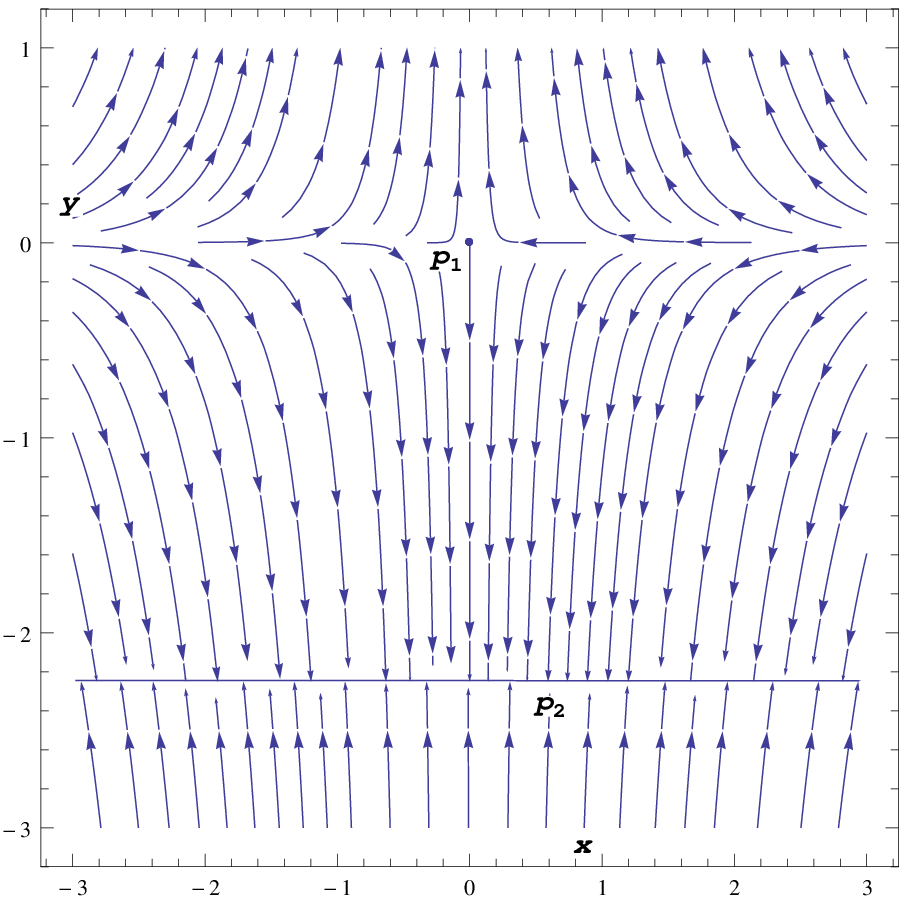}
\caption{Phase plot of the system when $\lambda = -10$ and $C=2$.}
\label{fig:minipage2}
\end{minipage}
\end{figure}

\par It deserves mention that $\lambda$ is not actually a constant. So the figures 1 and 2 represents a section of the 3-dimensional figure at particular values of $\lambda$, i.e., snapshots at those values. If one changes the values of $\lambda$, the nature of the figures remain the same. For a particular value of $\lambda$, equations (\ref{x}) and (\ref{y}) lead to an expression for the deceleration parameter $q$ in terms of scale factor $a$ as,\\

$q = -1 + \frac{3}{2} (1 - \frac{C^2}{3(1-C^2)} \frac{b a^{-\frac{3}{2} (\lambda - 1)}}{1 + \frac{b C^2}{3(1-C^2)} a^{-\frac{3}{2} (\lambda - 1)} })$.\\

Where $b$ is integration constant. In figure 3 and 4, the evolution of the deceleration parameter $q$ against $\frac{a}{a_{0}}$ is shown where $a_{0}$ is the present value of the scale factor. It is clearly seen that $\lambda <1$ is favoured in order to describe the observed dynamics of the universe. For $\lambda >1$, the universe starts from a negative $q$ ($q=-1$) with $x=0$ i.e., $\rho_{m}=0$ and settles into the stable configuration of a decelerated expansion. For $\lambda<1$, however, one obtains the desired behaviour, the universe starts with a deceleration ($q=\frac{1}{2}$) and settles into an accelerated phase with $q=-1$ and an arbitrary $\rho_{m}$. It should be noted that, for, $\lambda <1$, the interaction rate $\Gamma$ decays at a slower rate than the decay of $H$ and for a negative $\lambda$, the rate actually grows with some power of the decay of $H$.
 
\begin{figure}[H]
\centering
\begin{minipage}[b]{0.4\linewidth}
\includegraphics[scale=.6]{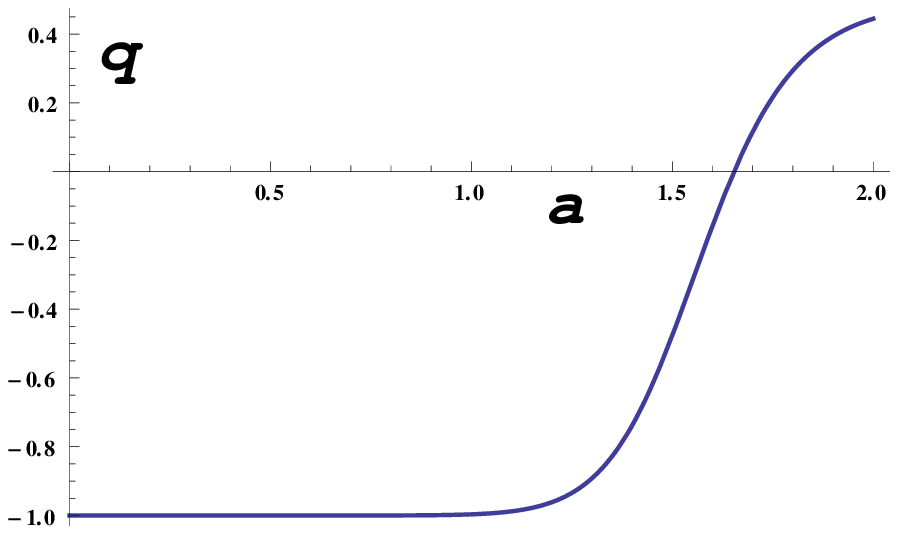}
\caption{$q$ vs $a$, when $\lambda = 10, C=2$ and $ b = -1000$.}
\label{fig:minipage1}
\end{minipage}
\quad
\begin{minipage}[b]{0.4\linewidth}
\includegraphics[scale=.5]{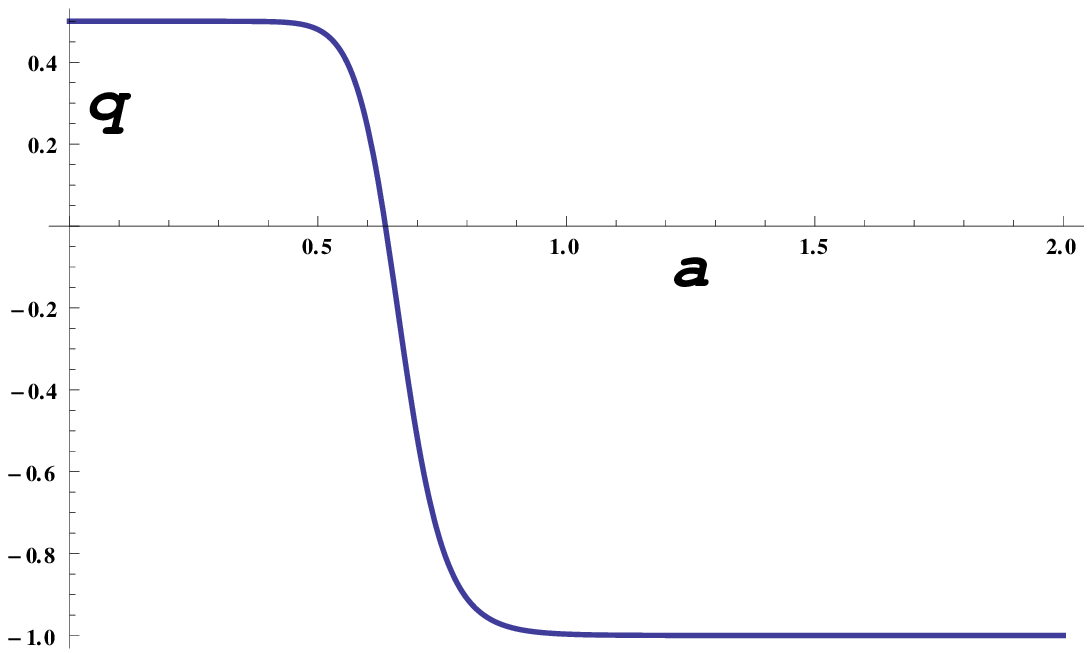}
\caption{$q$ vs $a$, when $\lambda = -10, C=2$ and $ b = -1000$.}
\label{fig:minipage2}
\end{minipage}
\end{figure}

\subsection{Class II : $\lambda =1$}
In this case $\Gamma$ is a linear function of $H$, given by $\Gamma = \alpha H$. Our system of equations reduces to 
\begin{equation} \label{x1}
x^{\prime} = - 3 x + \frac{c^2}{1-c^2} x y, 
\end{equation}
\begin{equation} \label{y1}
y^{\prime} = 0
\end{equation}

As $y$ is a constant, this is essentially a one dimensional system with $x=0$ and $y=\alpha$, a constant. Integration of the equation (\ref{x1}) yields the solution for $x$ as $x = A e^{-3 k N}$ where A is a constant of integration and $k= (1-\frac{C^2}{3(1-C^2)}\alpha)$. If $k>0$, the solution is indeed stable, as for $N \longrightarrow \infty$, one has $x \longrightarrow 0$. For $k<0$, the fixed point is unstable. The phase plot is shown in figure 5, which indicateds that the $y$-axis is the attractor of all solutions for $k>0$. Since $y$ is a constant, equation (\ref{q}) indicates that there is no transition from a decelerated to an accelerated expansion for the universe.

\begin{figure}[H]
\centering
\includegraphics[scale=.5]{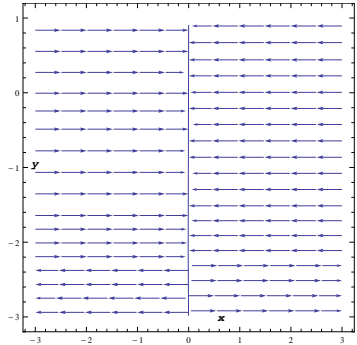}
\caption{Phase plot of the system when $\lambda = 1$ and $C=2$.}
\end{figure}

\section{Bifurcation in the system:}
It is interesting, from the point of view of the dynamical systems, to note that there is a clear bifurcation in the system, where $\lambda$ is the bifurcation parameter and $\lambda =1$ is the bifurcation point. When $\lambda<1$, there are two fixed points, $p_1$ and $p_2$ where $p_1$ is unstable but $p_2$ is stable. As $\lambda$ approaches unity these two merge into a single fixed point where the stability depends on the choice of the values of the constants $C$ and $\alpha$. With $\lambda>1$, i.e., when it attains values on the other side of the bifurcation point, one has the same two fixed points with their roles interchanged so far as the stability is concerned. Figures 1,2 and figure 3 show the change of the behaviour of the phase space with variation of $\lambda$, which clearly indicate the occurrence  of the bifurcation in the system. \\
At the bifurcation point, $\lambda =1$, $y=\frac{\Gamma}{H}$ is a constant. As already mentioned,  there is no transition from a decelerated to an accelerated expansion for the universe for a spatially flat FRW metric in this case. This result is completely consistent with that obtained by Pavon and Zimdahl\cite{diego2}.

\section{Discussion}
The holographic dark energy, tipped by many as the possible saviour from the coincidence problem, is analyzed as an autonomous system. It is found that the system indeed has unstable and stable fixed points. It is found that for $\lambda <1$, where $\lambda = (\dfrac{d\Gamma}{dH}) / (\frac{\Gamma}{H})$, the system indeed has at least one natural description of the universe  which starts with a decelerated expansion ($q=\frac{1}{2}$) and settles down to an accelerated expansion for the universe with dark matter completely giving way to the dark energy. So the interacting holographic dark energy warrants more attention, particularly when $\dfrac{d\Gamma}{dH} < \frac{\Gamma}{H}$ leading to $\lambda <1$. We see that the  interaction rate $\Gamma$ actually plays a crucial role in this. 

{\bf Acknowledgement:} N.R. wishes to thank CSIR (India) for financial support.


\begin{thebibliography}{99}
\bibitem{'thooft} G. 't Hooft, arxiv: gr-qc/9310026.
\bibitem{susskind} L. Susskind, J. Math. Phys., {\bf 36}, 6377 (1995).
\bibitem{diego} D. Pavon, J. Phys. A:Math. Theor., {\bf 40}, 6865 (2007).
\bibitem{li} M.Li, Phys. Lett. B {\bf 603}, 1 (2004).
\bibitem{diego1} D.Pavon and W. Zimdahl, Phys. Lett. B {\bf 628}, 206 (2005).
\bibitem{diego2} D.Pavon and W. Zimdahl, Class. Quantum Grav. {\bf 24}, 5461 (2007).
\bibitem{diego3} N. Banerjee and D. Pavon, Phy. Lett. B {\bf 647}, 477 (2007).
\bibitem{setare} M.R. Setare and M. Jamil, Phys. Lett B {\bf 690}, 1 (2010).
\bibitem{ellis} J. Wainwright and G. F. R. Ellis, \textit{Dynamical System in Cosmology} (Cambridge University Press, 2005).
\bibitem{coley} A.A.Coley, \textit{Dynamical System and Cosmology} (Springer, 2003).
\bibitem{gunzig} E. Gunzig, V. Faraoni, A. Figeredo and L. Brenig, Class. Quantum Grav. {\bf 17}, 1783 (2000).
\bibitem{carot} J. Carot and M.M. Collinge, Class. Quantum Grav. {\bf 20}, 707 (2003).
\bibitem{urena} L.A. Urena-Lopez, JCAP {\bf 0509}, 013 (2005).
\bibitem{nandan2} N. Roy and N. Banerjee, Eur. Phys. J. Plus, {\bf 129}, 162 (2014).
\bibitem{anjan} S. Kumar, S. Panda and A.A. Sen, Quantum Grav. {\bf 30}, 155011 (2013).
\bibitem{soma} S. Sen, A.A. Sen and M. Sami, Phys. Lett B. {\bf 686}, 1 (2010).
\bibitem{nandan1} N. Roy and N. Banerjee, Gen. Rel. Grav. {\bf 46}, 1651 (2014).
\bibitem{fang} W. Fang, H. Tu, J. Huang and C. Shu, arxiv:[1402.4005].
\bibitem{nairi1} N. Mazumder, R. Biswas and S. Chakraborty, arxiv:[1106.4627].
\bibitem{nairi2} N. Mazumder, R. Biswas and S. Chakraborty, arxiv:[1106.4626].
\bibitem{setare1} M.R. Setare and E.C. Vagenas, Int. J. Mod. Phys. D, {\bf 18}, 147 (2009).
\bibitem{strog} S.H. Strogatz, \textit{Nonlinear Dynamics and Chaos: With Applications to Physics, Biology, Chemistry and Engineering}; Westview Press, Boulder (2001). 
\bibitem{reza} R. Tavakol, `` Introduction to dynamical systems '' in ref \cite{ellis} .
\end{thebibliography}
 \end{document}